\pdfoutput=1
\documentclass[11pt]{article}
\usepackage{amsmath,amscd,amssymb,amsthm}
\usepackage{fourier}
\usepackage{graphicx}
\usepackage[sort&compress, comma, square, numbers]{natbib}
\usepackage[small, bf]{caption}
\usepackage{paralist}
\usepackage{multirow}
\usepackage{array}
\usepackage{rotating}
\usepackage{authblk}
\usepackage{enumitem}
\usepackage[colorlinks,linkcolor=black,bookmarksopen,
bookmarksnumbered,citecolor=black,urlcolor=black]{hyperref}
\usepackage[import]{xy}
\usepackage{color}

\graphicspath{{figs/harmonic/}}

\def\R{\mathbb{R}}

\def\F{\mathbb{F}}

\DeclareMathOperator{\im}{im}

\def\d{\operatorname{d}}

\DeclareMathOperator{\codiff}{\delta} 

\DeclareMathOperator{\dd}{d} 
\DeclareMathOperator{\hodge}{\ast} 

\DeclareMathOperator{\laplacian}{\Delta}

\newcommand{\norm}[1]{\lVert#1\rVert}

\newcommand{\pinnerproduct}[2]{( #1, #2 )}
\newcommand{\pInnerproduct}[2]{\bigl( #1, #2 \bigr)}

\long\def\symbolfootnote[#1]#2{\begingroup%
\def\thefootnote{\fnsymbol{footnote}}\footnote[#1]{#2}\endgroup}

\newtheorem*{theorem*}{Theorem}

\newtheorem*{proposition*}{Proposition}

\newtheorem*{lemma*}{Lemma}

\newtheorem*{claim*}{Claim}

\newtheorem*{axiom*}{Axiom}

\newtheorem*{conjecture*}{Conjecture}

\newtheorem*{corollary*}{Corollary}

\theoremstyle{definition}

\newtheorem*{definition*}{Definition}

\newtheorem*{example*}{Example}

\newtheorem*{exercise*}{Exercise}
\newtheorem*{recall*}{Recall}

\theoremstyle{remark}

\newtheorem*{note*}{Note}

\newtheorem*{remark*}{Remark}

\newtheorem*{notation*}{Notation}

\newtheorem*{question*}{Question}

\newtheorem*{fact*}{Fact}

\theoremstyle{theorem}
\newtheorem{theoremgi}{Theorem}

\theoremstyle{definition}

\theoremstyle{remark}

\DeclareMathOperator{\normal}{\textbf{n}}
\DeclareMathOperator{\tangential}{\textbf{t}}

\def\imagetop#1{\vtop{\null\hbox{#1}}}

\newenvironment{xyoverpic}[3]
{%
\begin{xy}
\xyimport#1{\includegraphics[#2]{#3}}
}{\end{xy}}

\allowdisplaybreaks

\begin{document}

\title{\vspace*{-1.4cm} Cohomologous Harmonic Cochains \footnote{This
    is a much shorter incarnation of version 6 of this paper which is
    available on arXiv as~\cite{HiKaWaWa2011v6}.}}  
\author{Anil N.~Hirani\thanks{Author
    for correspondence :
    \href{mailto:hirani@cs.illinois.edu}{hirani@cs.illinois.edu};
    \href{http://www.cs.illinois.edu/hirani}
    {http://www.cs.illinois.edu/hirani}}} \author{~Kaushik
  Kalyanaraman} \author{Han Wang} \author{Seth Watts}
\affil{University of Illinois at Urbana-Champaign}

\date{}

\maketitle

\begin{abstract}
  We describe algorithms for finding harmonic cochains, an essential
  ingredient for solving elliptic partial differential equations using
  finite element or discrete exterior calculus. Harmonic cochains are
  also useful in computational topology and computer graphics. We
  focus on finding harmonic cochains cohomologous to a given
  cocycle. Amongst other things, this allows for localization near
  topological features of interest. We derive a weighted least squares
  method by proving a discrete Hodge-deRham theorem on the isomorphism
  between the space of harmonic cochains and cohomology. The solution
  obtained either satisfies the Whitney form finite element exterior
  calculus equations or the discrete exterior calculus equations for
  harmonic cochains, depending on the discrete Hodge star used.
  \medskip

  \smallskip

  \noindent{\bf MSC Classes: } 65F10, 68U05, 65N30, 55-04; 
  {\bf ACM Classes: } F.2.2, G.1.6

\end{abstract}

\section{Introduction}

We discuss methods for finding simplicial harmonic cochains --
approximations of harmonic forms on simplicial meshes. In particular,
we want to find the harmonic cochain cohomologous to a given cocycle.
That is, given a cocycle $\omega$, we want a harmonic cochain $h$ such
that $h = \omega + \d \alpha$ for some $\alpha$. We either solve an
eigenvector problem followed by post processing or use a weighted least
squares method.

Harmonic cochains are used in finite element solution of elliptic
partial differential equations like the Poisson's equation
$\laplacian_p u=f$. See for instance \cite{ArFaWi2010}. They are also
useful in computer graphics for design of vector fields, since they
can provide a background on which vortices, sources and sinks may be
superimposed~\cite{FiScDeHo2007}. In computer graphics they are also
useful for finding conformal parameterization for texture mapping and
other applications~\cite{GuYa2008}.

We prove an easy discrete version of the Hodge-deRham isomorphism
theorem. This leads to a weighted least squares based method which is
the main contribution of this paper. The linear system is an obvious
one and can be derived also from the gradient part of Hodge
decomposition or in other ways.  The two other methods we describe are
based on finding eigenvectors followed by post processing. The least
squares method solves the mixed finite element exterior calculus
equations for harmonic cochains given in~\cite[Lemma
3.10]{ArFaWi2010}. (This is a result of Demlow and Hirani, and the
proof can be found in~\cite{HiKaWaWa2011v6}.) For each of the harmonic
cochain methods considered, the choice of the Hodge star operator
(Whitney or primal-dual) can be made, leading to two variations of
each method.

Other methods are those by Gu and Yau~\cite{GuYa2008}, and Desbrun et
al.~\cite{DeKaTo2008}. Both of these have some numerical disadvantages
especially when Whitney Hodge star is used instead of the diagonal
primal-dual Hodge star of discrete exterior calculus. (The Whitney
Hodge star is needed for general simplicial meshes, and for the lowest
order finite element exterior calculus.) In cases such as
2-dimensional cochains in tetrahedral meshes, the Desbrun et
al.~method does more work than is necessary for forming the linear
system, no matter which Hodge star is used.

\section{Preliminaries} \label{sec:prlmnrs}

Most of the needed background information on algebraic topology and
exterior calculus can be found in an earlier longer version of this
paper which is still available on arXiv~\cite{HiKaWaWa2011v6}. We use
two types of discretizations of exterior calculus -- discrete exterior
calculus, and finite element exterior calculus. In finite element
exterior calculus, we only consider the version that uses Whitney
forms.

We first recall the smooth Hodge-deRham theorem on the isomorphism
between cohomology and \emph{harmonic forms} ($\ker \laplacian)$ or
\emph{harmonic fields} ($\ker \d \cap \ker \codiff$). (This material
is based on~\cite{Schwarz1995}). The space of harmonic $p$-dimensional
fields on a manifold $M$ is denoted $\mathcal{H}^p(M)$. For a
\emph{closed manifold} (i.e., compact manifold without boundary),
harmonic forms and harmonic fields are the same, i.e., $\ker
\laplacian = \ker \d \cap \ker \codiff$. However, in the case of
compact manifolds with boundary $\partial M$, which we will refer to
as $\partial$-manifolds, one only has that $\ker \d \cap \ker \codiff
\subset \ker \laplacian$ and there can exist harmonic forms which are
not harmonic fields~\cite{CaDeGlMi2005}.

One of the striking properties of harmonic forms or fields is the link
they yield between topology and analysis or geometry. For closed
manifolds there is an isomorphism between real cohomology and the
space of harmonic forms. For compact $\partial$-manifolds however,
even the space of harmonic fields is infinite dimensional due to the
possibility of specifying boundary conditions. An isomorphism with
cohomology can be obtained by restricting harmonic fields by
specifying certain boundary conditions.

The \emph{tangential} component of a $p$-form $\omega$ is denoted
$\tangential \omega$ and its value is the value of $\omega$ on the
tangential (to $\partial M$) components of its vector field
arguments. Then the \emph{normal} component of $\omega$ is $\normal
\omega = \omega\vert_{\partial M} - \tangential
\omega$. See~\cite[page 27]{Schwarz1995} or~\cite[page
540]{AbMaRa1988}. These can also be defined using the pullback via the
inclusion map of the boundary into the manifold. A differential form
$\omega$ is said to satisfy the \emph{Neumann} or \emph{absolute}
boundary conditions if it has zero normal component ($\normal \omega =
0$), and the \emph{Dirichlet} or \emph{relative} boundary conditions
if it has zero tangential component ($\tangential \omega = 0$). Let
$\mathcal{H}^p_N(M)$ and $\mathcal{H}^p_D(M)$ be harmonic fields
satisfying the Neumann or Dirichlet boundary conditions,
respectively. Then one has:

\begin{theorem*}[Hodge-deRham
  Isomorphism~\cite{Schwarz1995}] \label{thm:smthhdgdrhm} If $M$ is a
  closed manifold, then $H^p(M; \R) \cong \mathcal{H}^p(M) = \ker
  \laplacian_p$, and if it is a compact $\partial$-manifold then
  $H^p(M;\R)\cong \mathcal{H}^p_N(M)$ and $H^p(M, \partial M; \R)
  \cong \mathcal{H}^p_D(M)$.
\end{theorem*}

The space $H^p(M; \R)$ is the (absolute) real $p$-cohomology vector
space of $M$, and $H^p(M,\partial M; \R)$ is the relative real
$p$-cohomology vector space of $M$, relative to its boundary. For
$\partial$-manifolds, we will only consider harmonic fields satisfying
Neumann conditions. This is because the least squares method is based
on a weak form of the Laplace-deRham operator, and in that framework
the Neumann conditions are automatic, that is they do not have to be
enforced explicitly. For manifold complexes with boundary we will use
\emph{harmonic cochains} synonymously with \emph{harmonic Neumann
  cochains}.

\section{Eigenvector Methods} \label{sec:eigen}

Cohomologous harmonic cochains can be computed by first computing a
harmonic cochain basis followed by some post processing. Such a basis
can be obtained as eigenvectors of the zero eigenvalue of a discrete
$\laplacian_p$. The problem of finding eigenvectors can be formulated
(in the terminology of finite element methods) using a weak mixed or
weak direct method. While nothing is published about the eigenvector
method, the weak mixed method was the one used by Arnold et
al.~\cite{ArFaWi2010} in one of their examples.

Let $\laplacian = \d \codiff + \codiff \d$ be the smooth
Laplace-deRham operator on some manifold $M$. Then the direct
eigenvalue problem is to find a nonzero differential form $u$ and a
real scalar $\lambda$ such that $\laplacian u = \lambda u$.  The
formal derivation of the weak direct method goes like this: start by
posing the problem of finding a $u$ such that
$\pinnerproduct{\laplacian u}{v} = \lambda \pinnerproduct{u}{v}$ for
all $v$, the inner products being those on forms. Then using the
formula for the Laplace-deRham operator, and assuming appropriate
boundary conditions (which implies adjointness of $\d$ and $\codiff$)
this is equivalent to finding a $u$ such that $\pinnerproduct{\d u}{\d
  v}+ \pinnerproduct{\codiff u}{\codiff v} = \lambda
\pinnerproduct{u}{v}$ for all $v$.  If $M$ is replaced by its
simplicial complex approximation (which we will also refer to as $M$)
then the discretization yields the linear system $\laplacian_p u =
\lambda \hodge_p u$, where now $\laplacian_p := \d_p \hodge_{p+1} \d_p
+ (-1)^{(p-1)(n-p+1)} \hodge_p \d_{p-1} \hodge^{-1}_{p-1} \d^T_{p-1}
\hodge_p$ is the discrete Laplace-deRham
operator~\cite{HiKaWaWa2011v6} and $u$ is a $p$-cochain. Here
$\hodge_p$ is the mass matrix for Whitney $p$-forms or the primal-dual
discrete Hodge star. The harmonic cochains are thus the solutions
corresponding to the zero eigenvalue for this generalized eigenvalue
problem.

For the weak mixed eigenvector method, consider the linear system for
the unknowns $\sigma$ and $u$:
\begin{align*}
  \pinnerproduct{\sigma}{\tau}-\pinnerproduct{\d_{p-1}\tau}{u} &=0\,,\\
  \pinnerproduct{\d_{p-1}\sigma}{v} + \pinnerproduct{\d_p u}{\d_p v}&=0\,,
\end{align*}
for all $\tau$ and $v$. Then $(\sigma, u)$ is a solution if and only
if $\sigma = 0$ and $u$ is a harmonic $p$-form~\cite[Lemma
3.10]{ArFaWi2010}. We discretize these equations and obtain the system
matrix
\begin{equation}\label{eq:mxdmtrx}
\begin{bmatrix}
  -\hodge_{p-1} & \d_{p-1}^T \hodge_p\\
  \hodge_p \d_{p-1} & \d_p^T\hodge_{p+1} \d_p
\end{bmatrix} \, ,
\end{equation}
whose eigenvectors corresponding to the zero eigenvalue we seek.

\begin{figure}[p]
\centering
\begin{tabular}{c}
  \begin{tabular}{cc}
    \includegraphics[scale=0.30, trim=3.1in 2.3in 3.1in 3.0in, clip]
    {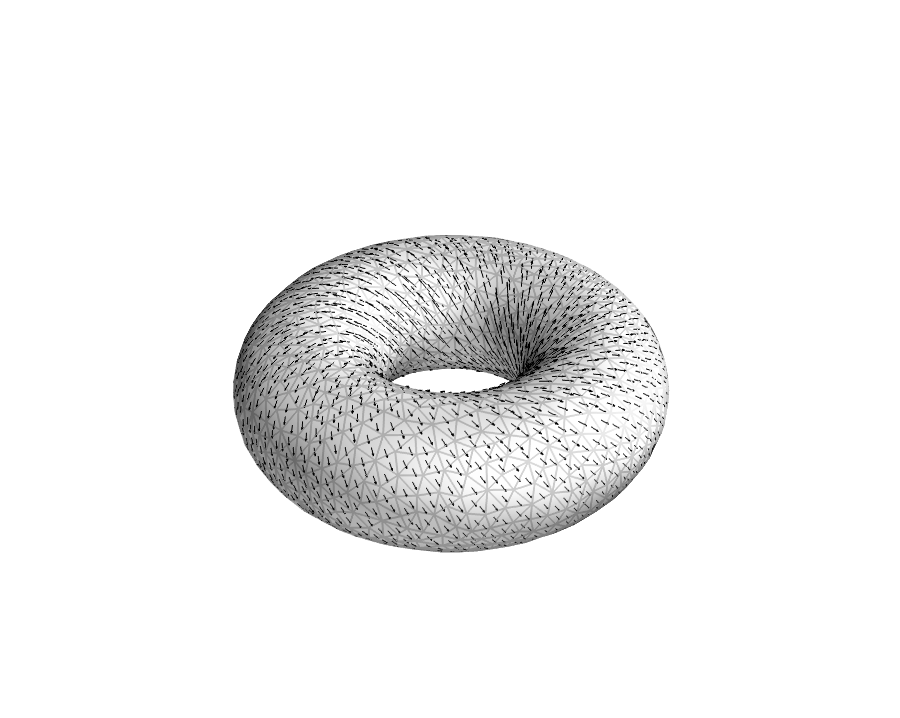} & 
    \includegraphics[scale=0.30, trim=3.1in 2.3in 3.1in 3.0in, clip]
    {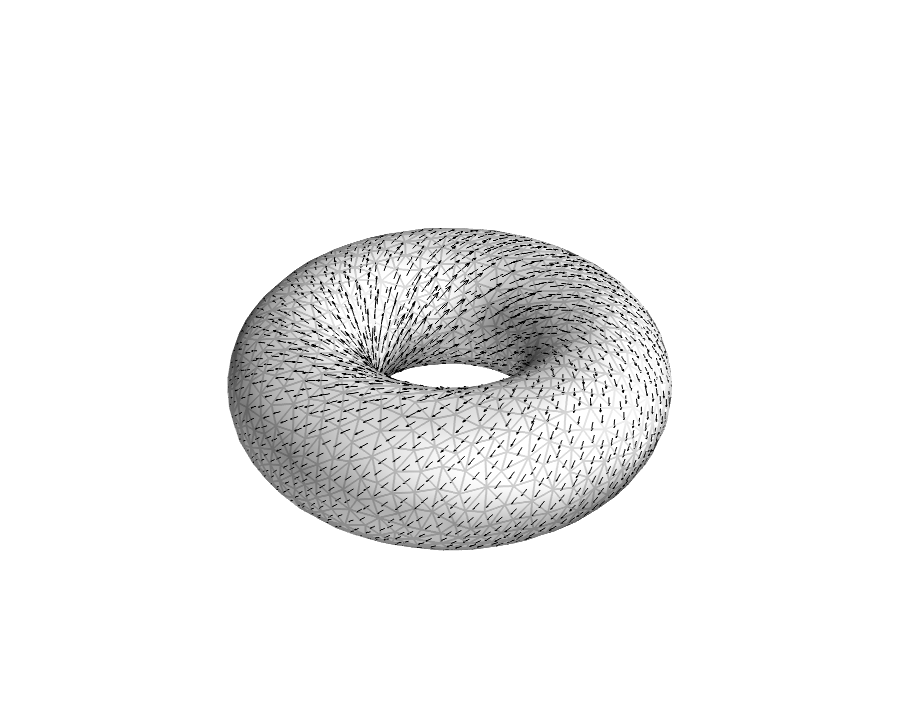} 
  \end{tabular} \\
  \begin{tabular}{cc}
    \includegraphics[scale=0.35, trim=2.0in 0.9in 1.8in 0.9in, clip]
    {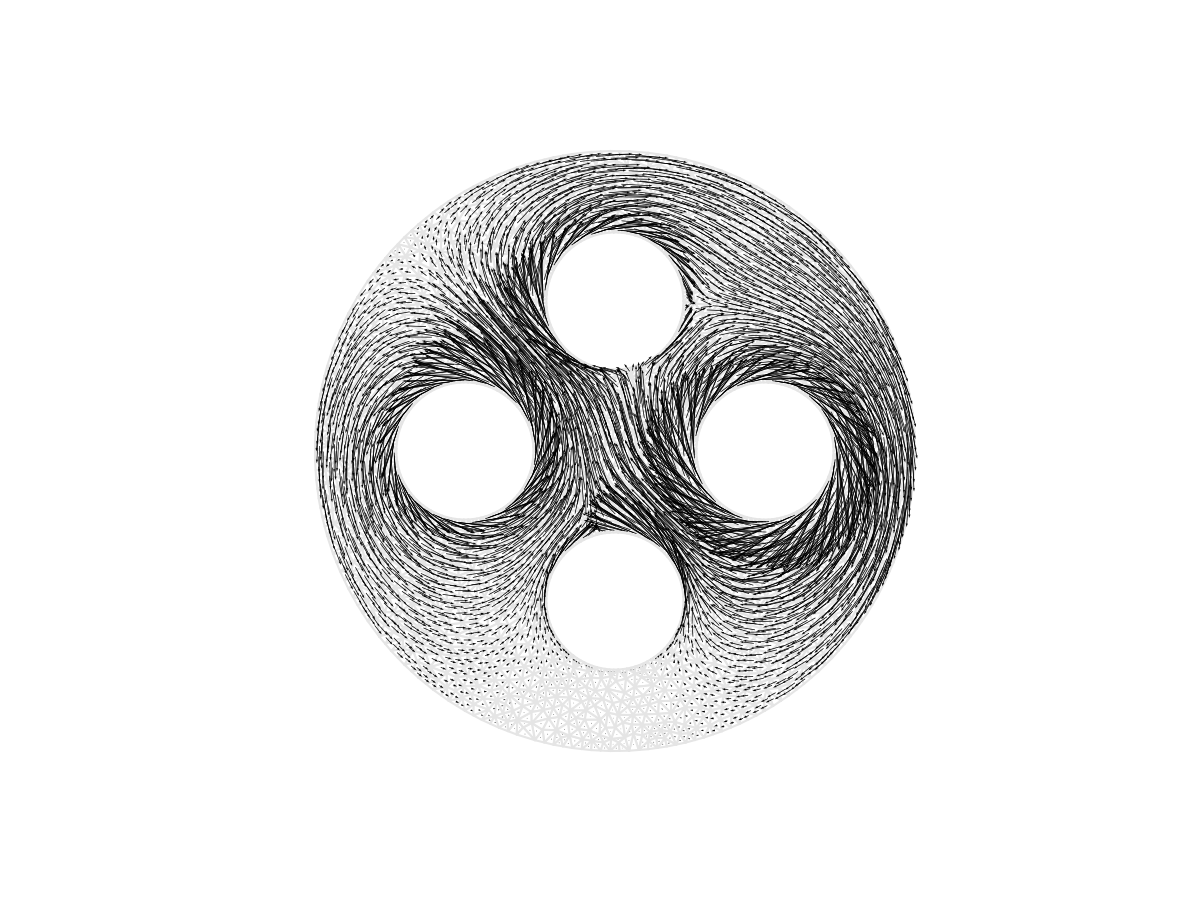} &
    \includegraphics[scale=0.35, trim=2.0in 0.9in 1.8in 0.9in, clip]
    {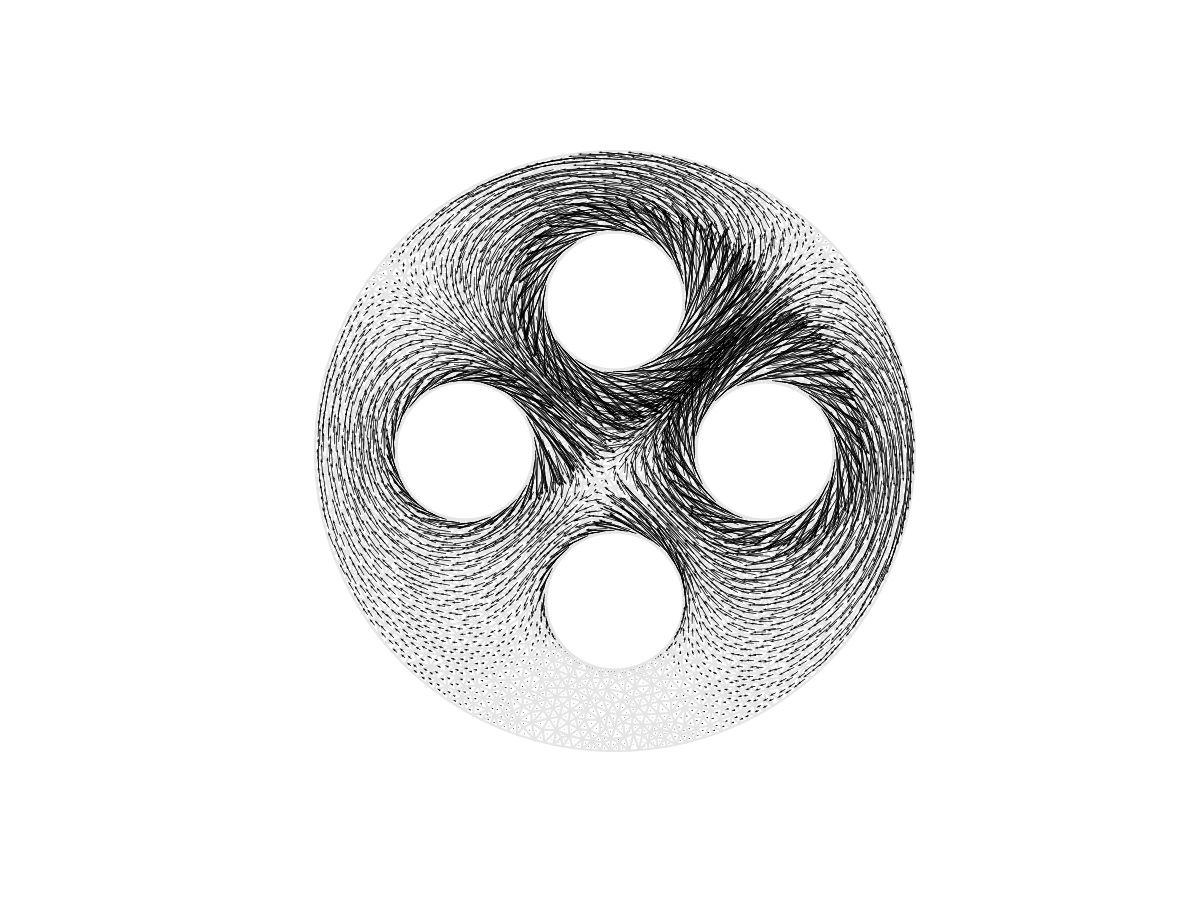} \\
    \includegraphics[scale=0.35, trim=2.0in 0.9in 1.8in 0.9in, clip]
    {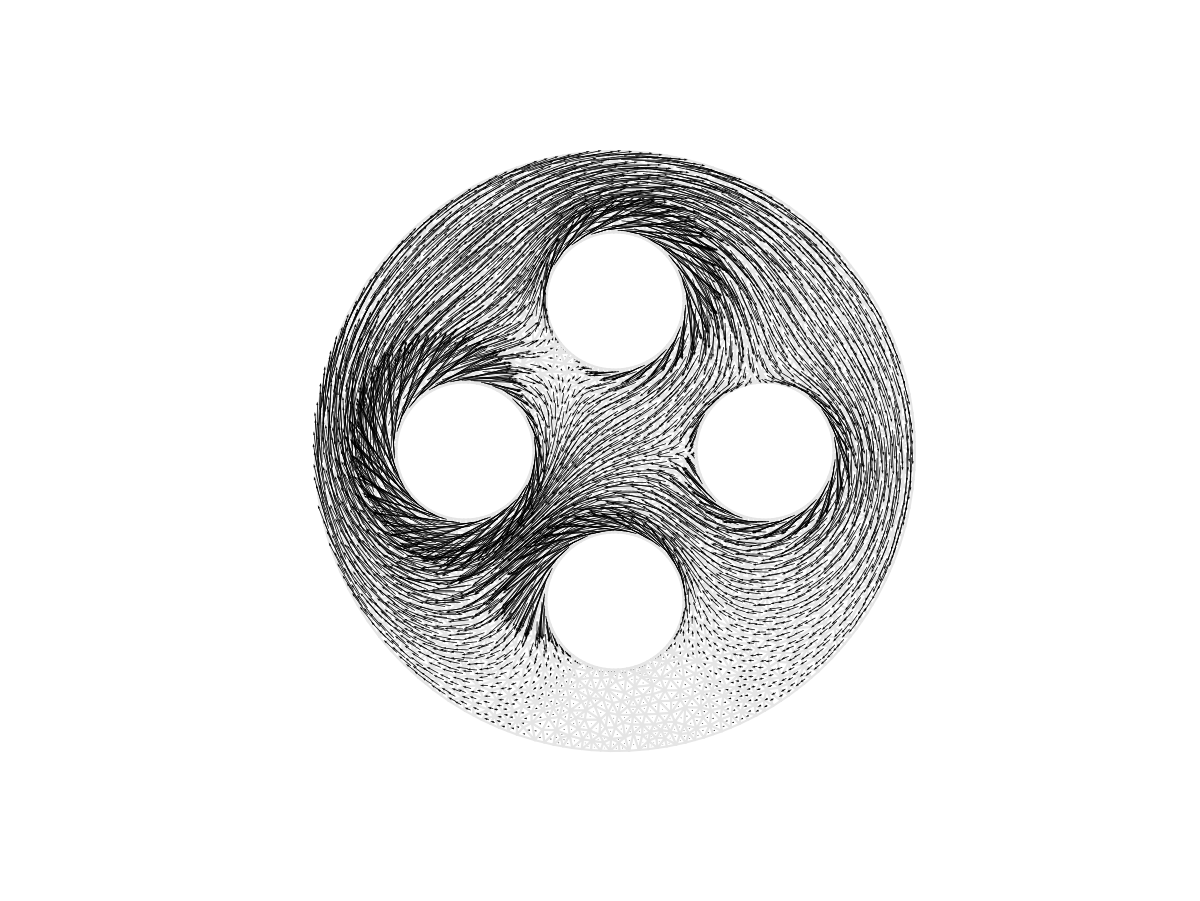} &
    \includegraphics[scale=0.35, trim=2.0in 0.9in 1.8in 0.9in, clip]
    {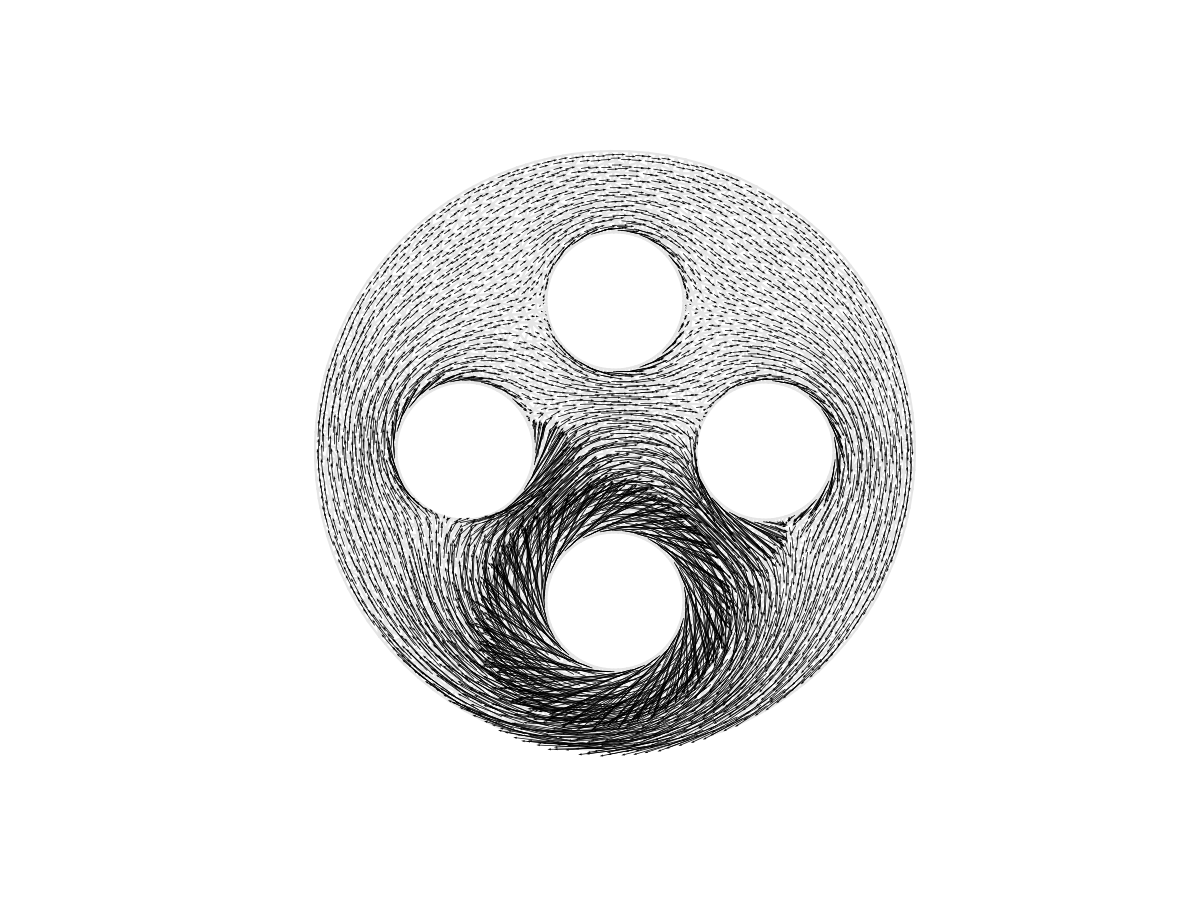}
  \end{tabular}
\end{tabular}
\caption{Harmonic cochains produced by the mixed eigenvector
  method. The torus has a two-dimensional space of harmonic cochains
  and the four-holed disc has a four-dimensional space of harmonic
  Neumann fields.}
\label{fig:eigen}
\end{figure}

Figure~\ref{fig:eigen} shows results of the eigenvector
calculations.The eigenvector methods will often suffice, if all that
is needed is \emph{some} harmonic basis, which may be the common case
in finite element exterior calculus. Applications like vector field
design in computer graphics may require more control over the process,
namely the satisfaction of the cohomology constraint.

\subsection{Projection based methods} \label{sub:prjctn}

If a harmonic cochain basis is available, then orthogonal projection
to the harmonics can be used to obtain a harmonic cochain $h$
cohomologous to a given cocycle $\omega$. (This method was suggested
to us by Ari Stern.) In contrast, the least squares method discussed
in Section~\ref{sec:lstsqrs} finds a cohomologous harmonic cochain
without requiring any precomputation of a harmonic basis. Moreover,
the projection method does not find the potential of the gradient
part. If that is needed, then the least squares method
equation~\eqref{eq:lstsqrs} has to be solved anyway.

Let $H$ be a matrix whose columns form a harmonic $p$-cochain basis.
Given a nontrivial cocycle $\omega$, we seek the harmonic cochain $h$
such that $h = \omega + \d \alpha$ for some $\alpha$. (Thus we are
interested in a Hodge decomposition of $\omega$. Note that the Hodge
decomposition of an arbitrary $\omega$ would be $\d \alpha + \codiff
\beta + h$, but since the given $\omega$ is a nontrivial cocycle, it
has no curl part.)  Since $\im \d$ is orthogonal to every column $h_i$
of $H$, we have that $\pInnerproduct{\omega + \d \alpha}{h_i} =
\pInnerproduct{\omega}{h_i} = \left(\sum_j a_j h_j, h_i\right)$ for
all $i$, where $\sum_j a_j h_j$ is the $h$ that we seek. (The inner
product above is the $p$-cochain inner product.) Writing $a$ for the
vector of unknown coefficients $a_j$, we can express the last equality
as the linear system $H^T \!  \hodge H \, a = H^T \hodge
\omega$. After solving this for the unknowns $a$, the vector $H\, a$
is the desired $h$. This is the normal equation for a weighted least
squares problem (a different system from theh one in
Section~\ref{sec:lstsqrs}). The matrix of the linear system is of
order of the $p$-Betti number and the cost of this projection will be
dominated by the matrix vector multiplications needed in forming $H^T
\!  \hodge H$ if the Betti number is small. If the columns of $H$ are
orthonormal in the $\hodge$ inner product then no linear solve is
required.

\subsection{Pairing with homology basis}
\label{sub:pairing}

For vector field design in computer graphics or in physical
applications, the usual cases are dimension 2 with 1-cochains and
dimension 3 with 1-cochains or 2-cochains. In the latter case, only
solid handles and cavities are relevant since general 3-manifolds are
typically not used in such applications. In all these cases, it makes
sense to talk of homology basis elements corresponding to topological
features. These can be used by pairing with cohomology to find
cohomologous harmonic cochains. (This method was suggested to us by
Douglas Arnold.) For this one needs an explicit isomorphism $H^p(M)
\cong H_p(M)^\ast$, where $H_p(M)^\ast$ is the vector space dual of
real-valued homology. Note that this method requires not only the
entire harmonic cochain basis but also a full homology basis.

Given $[\omega]\in H^{p}(M)$, define the map $\varphi:
H^{p}(M)\rightarrow H_{p}(M)^{*}$ by $\varphi[\omega][z]:= \omega(z)$
for any $[z]\in H_{p}(M)$. This map is well-defined: given other
representatives $\omega + \d \alpha$ and $z + \partial y$, one has
$(\omega + \d \alpha) (z + \partial y) = \omega(z) + \d \alpha(z) +
\omega(\partial y) + \d \alpha(\partial y) = \omega(z)$. To prove that
$\varphi$ is an isomorphism, it is enough to show that it is
injective. That is, we would like to show that for any $[\omega]\in
H^{p}(M)$, $\omega(z) = 0$ for all $[z] \in H_{p}(M)$ implies that
$[\omega] = 0$. This is equivalent to showing that if $\omega$ is a
representative of an element of $H^{p}(M)$, $\omega(z) = 0$ for all
nontrivial cycles $z$ implies that $\omega$ is exact. Since $\omega$
is nontrivial, $\omega = \d \alpha + h$ for some $\alpha$ and harmonic
cochain $h$. To show $\omega$ is exact is the same as showing $h =
0$. Thus we have to show that given a harmonic cochain $h$, $h(z) = 0$
for all nontrivial cycles $z$ implies that $h = 0$. We now show this
for the case of $n = 2$ ($\dim $ of $M$) and $p = 1$.

\setcounter{theoremgi}{0}
\begin{theoremgi}
  Let $M$ be a surface simplicial complex. Then $\varphi : H^{1}(M)
  \rightarrow H_1(M)^{\ast}$ is injective (hence an isomorphism).
\end{theoremgi}

\begin{proof}
  It is enough to consider a homology basis of nontrivial
  cycles. Suppose $M$ has $b$ holes and $g$ handles. Consider a
  homology basis corresponding to the holes, handles and tunnels. That
  is, let $z_1 \, , \dots , z_{b - 1}$ be cycles corresponding to $b -
  1$ of the $b$ holes (the remaning one hole is considered the outer
  boundary), $\mu_1 \, , \dots , \mu_g$ be handle cycles corresponding
  to the $g$ handles, $\lambda_1 \, , \dots , \lambda_g$ be tunnel
  cycles corresponding to the $g$ handles. (Handle cycles are like
  longitudes on a torus and tunnel cycles are like latitudes on a
  torus.) Let $\omega_z$ be the collection of $b - 1$ nontrivial
  cocycles corresponding to the hole cycles in, and let $\omega_{\mu}$
  and $\omega_{\lambda}$ be similarly defined. Each such cocycle
  $\omega$ is a ``picket fence'' (see
  Figure~\ref{fig:lstsqrschmlgy}). Either two edges of a triangle
  carry a part of $\omega$ or none. The hole cocycles join the
  boundary of a hole to the outer boundary. The handle and tunnel
  coycles go around the handle or tunnel. Such cocycles are obtained
  by dualizing cycles on the dual mesh. By
  Theorem~\ref{thm:dscrthdgdrhm}, there is a basis of cohomologous
  cochains $h_z$ for all $z$, $h_{\mu}$ for all $\mu$, and
  $h_{\lambda}$ for all $\lambda$ (cohomologous to the corresponding
  $\omega$'s).

  Now consider a harmonic 1-cochain that evaluates to 0 on all the
  basis cycles above. In terms of the harmonic basis above,

  \begin{equation} \label{eq:hrmnccmbntn} 
    h = \sum\limits_z r_z h_z +
    \sum\limits_{\mu} r_{\mu} h_{\mu} + \sum\limits_{\lambda}
    r_{\lambda} h_{\lambda} \, .
  \end{equation}

  Note that $h_z$ evaluates to nonzero on $z$ and 0 on every other
  cycle, $h_{\mu_i}$ evaluates to nonzero on $\lambda_i$ and 0 on
  every other cycle, $h_{\lambda_i}$ evaluates to nonzero on $\mu_i$
  and 0 on every other cycle. $h_z (z) = \omega_z (z) = \omega_z
  (B_z)$, where $B_z$ is the boundary of the corresponding hole since
  $h_z$ is cohomologous to $\omega_z$ and $z$ is homologous to
  $B_z$. But $\omega_z (B_z) = \pm 1$ (or whatever value was picked
  for edges). Likewise, $h_z (z') = \omega_z (z') = \omega_z (B_{z'})
  = 0$, for $z' \neq z$ since $\omega_z$ takes value 0 on edges of
  $B_{z'}$. Similarly for $h_z$ on other types of cycles, and for the
  other harmonic basis elements. Thus, the coefficients
  in~\eqref{eq:hrmnccmbntn} are all zero.
\end{proof}

If $M$ is the closure of a connected open subset of $\R^3$ and the
topological features of interest are cavities and solid handles then a
result similar to the above one can be shown.  Now let $H$ be a matrix
whose columns form a basis of harmonic $p$-cochains and $B$ a matrix
whose columns form a homology basis corresponding to topological
features in the sense described in the proof above. Then $(B^{T}
H)^{-1}$ contains the harmonic cochains cohomologous to the
topological features.

\section{Least Squares Method} \label{sec:lstsqrs}

In what follows, $M$ will be a simplicial manifold complex, with or
without boundary. All references to $\laplacian$ are to the discrete
Laplace-deRham operators~\cite{HiKaWaWa2011v6}.  For a closed manifold,
one way to show the Hodge-deRham isomorphism theorem of
Section~\ref{sec:prlmnrs} for the smooth case is to use a variational
approach~\cite[Theorem 2.2.1]{Jost2005}. One shows that in each
cohomology class there is exactly one harmonic form and it is the one
with the smallest norm. The norm used is the $L^2$ norm induced from
the inner product of differential forms. Inspired by this, we
formulate a simple discrete version of this theorem. This is done for
harmonic cochains in the case of manifold simplicial complexes without
boundary, and for harmonic Neumann cochains in the case with
boundary. First we derive the necessary stationarity conditions in the
discrete case.  For $\omega\in C^{p}$ s.t. $\dd_p\omega=0$, we
consider the optimization problem $\min_{\alpha \, \in \, C^{p-1}}
\pinnerproduct{\omega + \d_{p-1}\alpha\,} {\,\omega +
  \dd_{p-1}\alpha}_{C^p}$, where the
$\pinnerproduct{\cdot}{\cdot}_{C^p}$ is the inner product on
$p$-cochains~\cite{BeHi2011a}. Writing this in matrix notation, we want
to find the minimizer $\alpha$ in the optimization problem

\begin{equation} \label{eq:min}
  \min_{\alpha\in C^{p-1}} \bigl(\omega + \dd_{p-1}
  \alpha\bigr)^T \hodge_p \, \bigl(\omega + \dd_{p-1}
  \alpha\bigr) \, .
\end{equation}
From the stationary condition for the minimizer and using properties
of the Hodge star matrix, we obtain:
\begin{equation} \label{eq:lstsqrs}
\dd_{p-1}^T \hodge_p \dd_{p-1} \alpha = -\dd_{p-1}^T \hodge_p 
\omega \, .
\end{equation}

This is the normal equation for the weighted least squares problem
$\d \alpha \simeq -\omega$. Although the above equation is a necessary
condition for solving the optimization problem~\eqref{eq:min}, the
matrix $\d_{p - 1}^T\hodge_p\d_{p - 1}$ may have a nontrivial
kernel. In fact in the interesting cases it generally will. (For
example, for $p=1$, the $\ker \d_0$ will have dimension equal to the
number of connected components in the complex.)  Thus, for $\alpha$ to
be a minimizer we need that the Hessian $\d_{p - 1}^T\hodge_p\d_{p -
  1}$ be at least positive semidefinite, which is true because of the
positive definiteness of $\hodge_p$. In this case, $\alpha$ may not be
unique, but as we will show next, $\d_{p-1}\alpha$ will be
unique. Note that equation~\eqref{eq:lstsqrs} is equivalent to
$\codiff_p \d_{p-1} \alpha = -\codiff_p \omega$ which is $\codiff_p
(\omega + \d_{p-1} \alpha) = 0$. This should make the connection to
$\omega + \d_{p-1} \alpha$ being harmonic more transparent since we
also have that $\d (\omega + \d \alpha) = 0$.

From the above, if $[\omega] \in H^p(K \, ; \R)$, then it is easy to
see that
\begin{inparaenum}[\itshape (i)]
\item there exists a cochain $\alpha \in C^{p - 1}(K \, ; \R)$, not
  necessarily unique, such that $\codiff_p \, (\omega + \d_{p - 1}
  \alpha) = 0$;
  \item there is a unique cochain $\d_{p - 1} \alpha$ satisfying
    $\codiff_p \, (\omega + \d_{p - 1} \alpha) = 0$ ; and
  \item $\codiff_p \, (\omega + \d_{p - 1} \alpha) = 0$ implies
    $\laplacian_p \, (\omega + \d_{p - 1} \alpha) = 0$.
  \end{inparaenum}
  To see
  \begin{inparaenum}[\itshape (i)]
  \item consider the least squares problem $\d_{p - 1} a \simeq
    \omega$. Let $-\alpha$ be a solution. Some such $\alpha$ always
    exists because least squares problems always have a solution. Note
    that the norm used in formulating this problem as a residual
    minimization is the one induced from the Hodge star inner product
    on cochains. Specifically, the inner product matrix is $\hodge_p$
    and the least squares problem minimizes $(\omega + \d_{p - 1}
    \alpha)^T \hodge_p (\omega + \d_{p - 1} \alpha)$ since $\omega -
    \d_{p - 1} \, (-\alpha) = (\omega + \d_{p - 1} \alpha)$ is the
    residual. But from properties of least squares~\cite{Bjork1996}
    the residual $(\omega + \d_{p - 1} \alpha)$ is
    $\hodge_p$-orthogonal to $\im \d_{p - 1}$. Thus we have that
    $(\omega + \d_{p - 1} \alpha) \in \im {\d_{p -
        1}}^{\perp_{\hodge_p}} = \ker \codiff_p$ since $\codiff_p$ is
    the adjoint of $\d_{p - 1}$ up to sign in the Hodge star inner
    product on cochains. In \item uniqueness of $\d_{p - 1} \alpha$
    follows from properties of least squares, and \item is obvious
    since $\omega + \d_{p-1} \alpha$ is also closed.
\end{inparaenum}
Note that unlike in the smooth case, $\codiff_{p+1}$ and $\d_p$ are
adjoints of each other up to sign only. Specifically,
$\pInnerproduct{\d_p \alpha}{\beta}_{C_{p+1}} \! \! = (-1)^{1 - p^2}
\pInnerproduct{\alpha}{\codiff_{p + 1} \beta}_{C_p}$ for any
$p$-cochain $\alpha$ and $(p+1)$-cochain $\beta$. From the preceding
discussion, we have the following elementary but useful theorem:

\begin{theoremgi}[Discrete Hodge-deRham
  Isomorphism] \label{thm:dscrthdgdrhm} There is a unique harmonic
  cochain in each cohomology class and it is the one with the smallest
  norm. Given a cocycle $\omega$ its cohomologous harmonic cochain is
  $\omega + \d \alpha$ where $\alpha$ is a solution of $\dd^T \hodge
  \dd \alpha = -\dd^T \hodge \omega$.
\end{theoremgi}

An alternative derivation of \eqref{eq:lstsqrs} is to project $\omega$
to image of $\d$ by requiring that $\pinnerproduct{\d \alpha}{\d \tau}
= \pinnerproduct{\omega}{\d \tau}$ for all $\tau$.  Yet another
derivation is the following. Given an $\omega$, to find its Hodge
decomposition, one starts with $\omega = \d \alpha + \codiff \beta +
h$, where we are seeking a harmonic field or cochain $h$ and an
$\alpha$ and $\beta$. Applying $\codiff$ to both sides yields
$\hodge^{-1} \d^T \! \hodge \d \alpha = \hodge^{-1} \d^T \! \hodge
\omega$, which is the same as~\eqref{eq:lstsqrs} up to sign after the
$\hodge^{-1}$ is cancelled from both sides. Note that the linear
system for the $\beta$ part is $\d \codiff \beta = \d \omega$. This
has a $\hodge^{-1}$ which cannot be removed by cancellation since this
is $\d \hodge^{-1} \d^T \! \hodge \beta = \d \omega$.  In his
thesis~\cite{Bell2008}, Bell was motivated by the need to address the
inverse Hodge star matrix in order to apply algebraic multigrid to the
Hodge decomposition problem. He proposed replacing the Hodge stars by
identity and solving the above systems starting with random cochains
until one has obtained a cohomology basis. (He did not prove that the
procedure is guaranteed to produce such a basis.) He then showed that
choosing the basis elements as $\omega$ and solving~\eqref{eq:lstsqrs}
for each one yields a basis of harmonic cochains. In contrast we have
shown above that each such $\omega$ is cohomologous to the
corresponding harmonic cochain $h$ individually.

\begin{figure}[p]
  \centering
  \begin{tabular}{c}
    \begin{tabular}{cc}
      \includegraphics[scale=0.3, trim=2.9in 2.2in 2.9in 2.5in, clip]
      {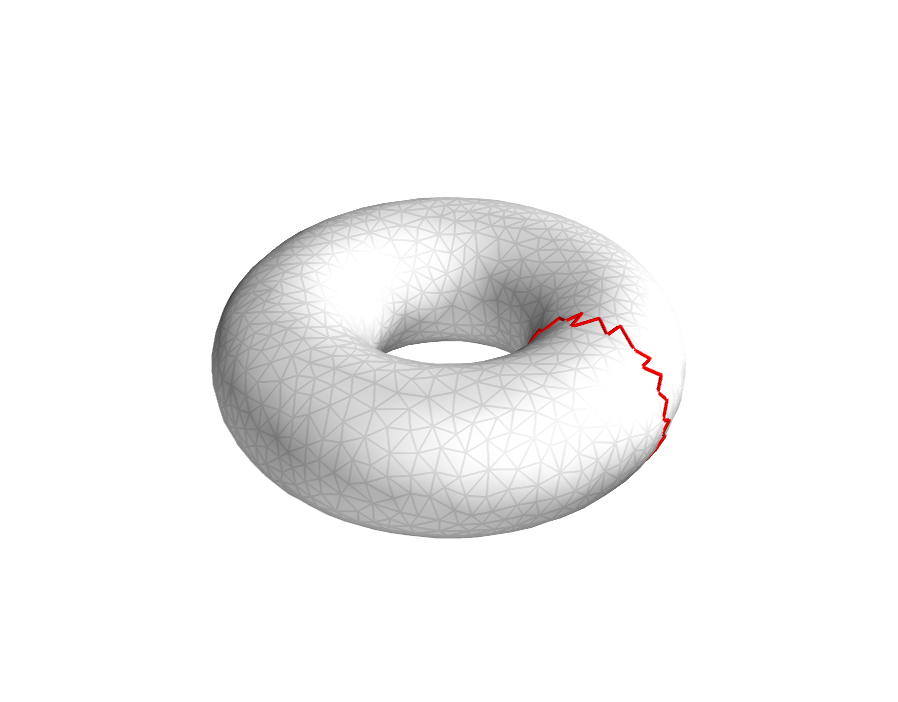} &
      \includegraphics[scale=0.3, trim=2.9in 2.2in 2.9in 2.5in, clip]
      {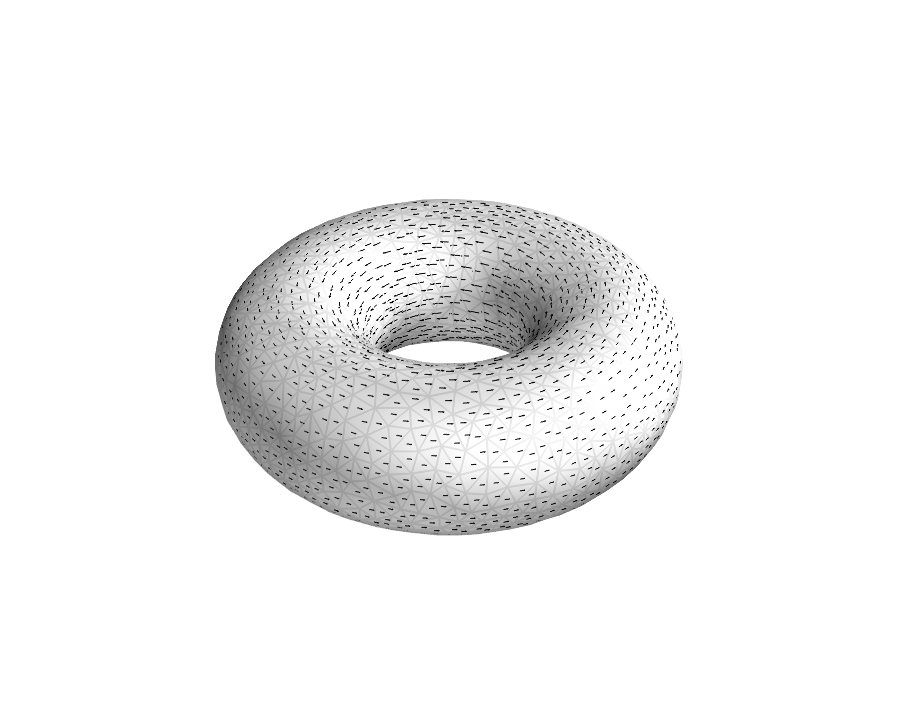} \\
      \includegraphics[scale=0.3, trim=2.9in 2.2in 2.9in 2.5in, clip]
      {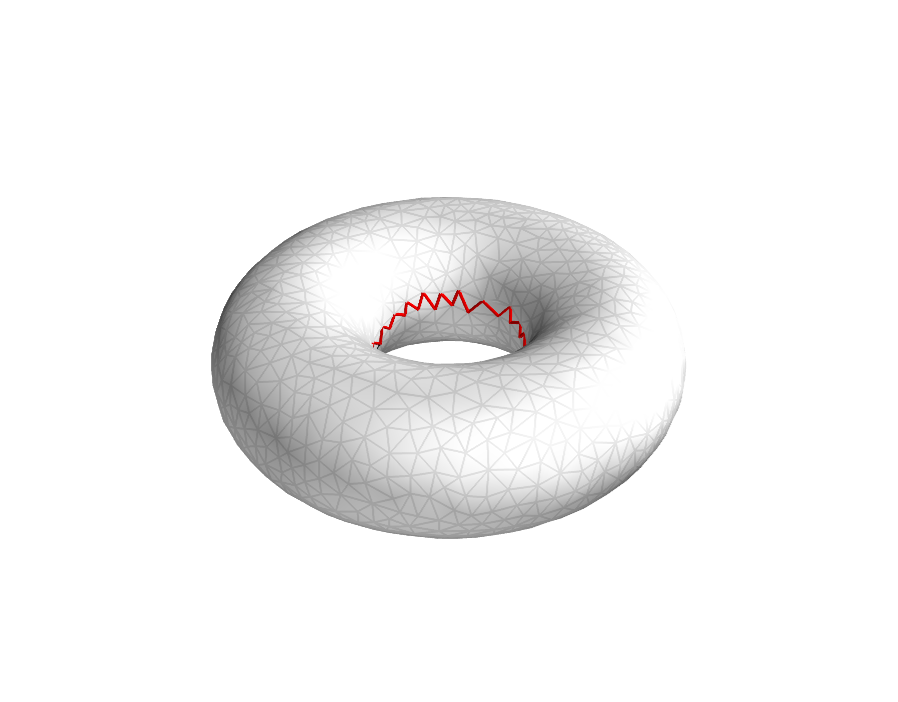} &
      \includegraphics[scale=0.3, trim=2.9in 2.2in 2.9in 2.5in, clip]
      {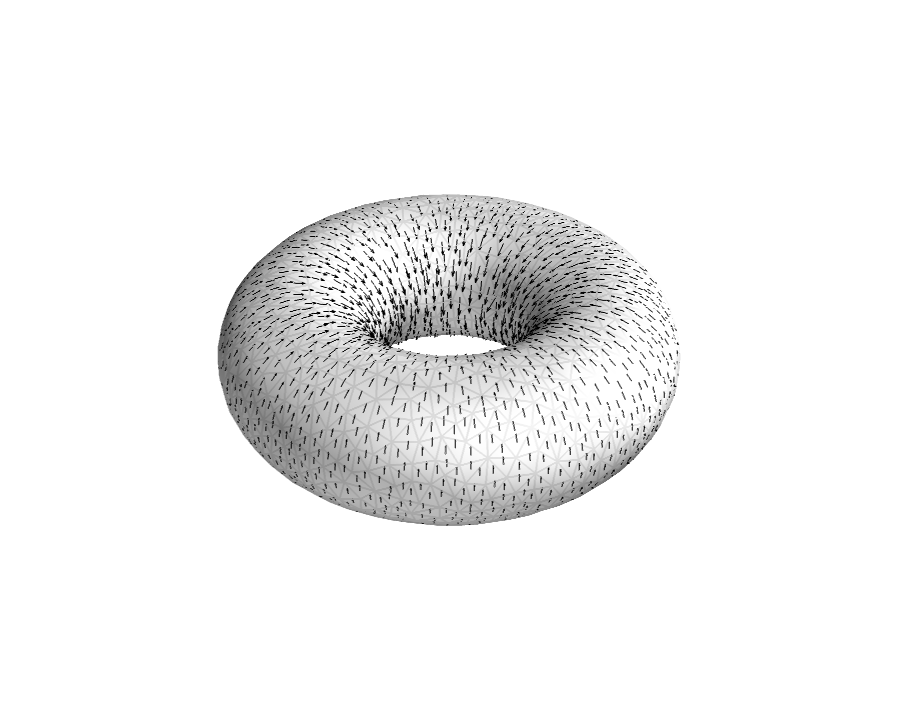} 
    \end{tabular} \\
   \begin{tabular}{ccc}
      \includegraphics[scale=0.35, trim=2.05in 0.95in 1.85in 0.95in,
      clip]
      {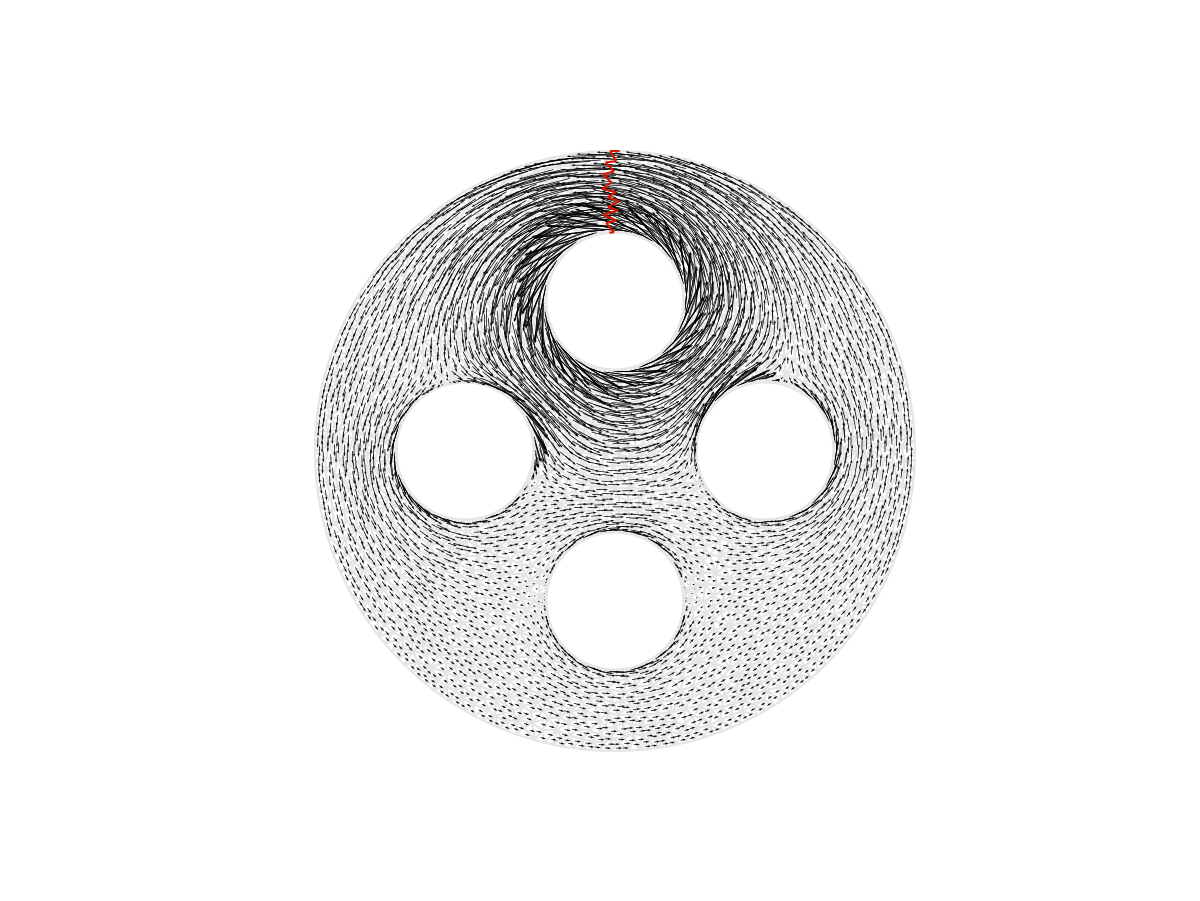} &
      \includegraphics[scale=0.35, trim=2.05in 0.95in 1.85in 0.95in,
      clip]
      {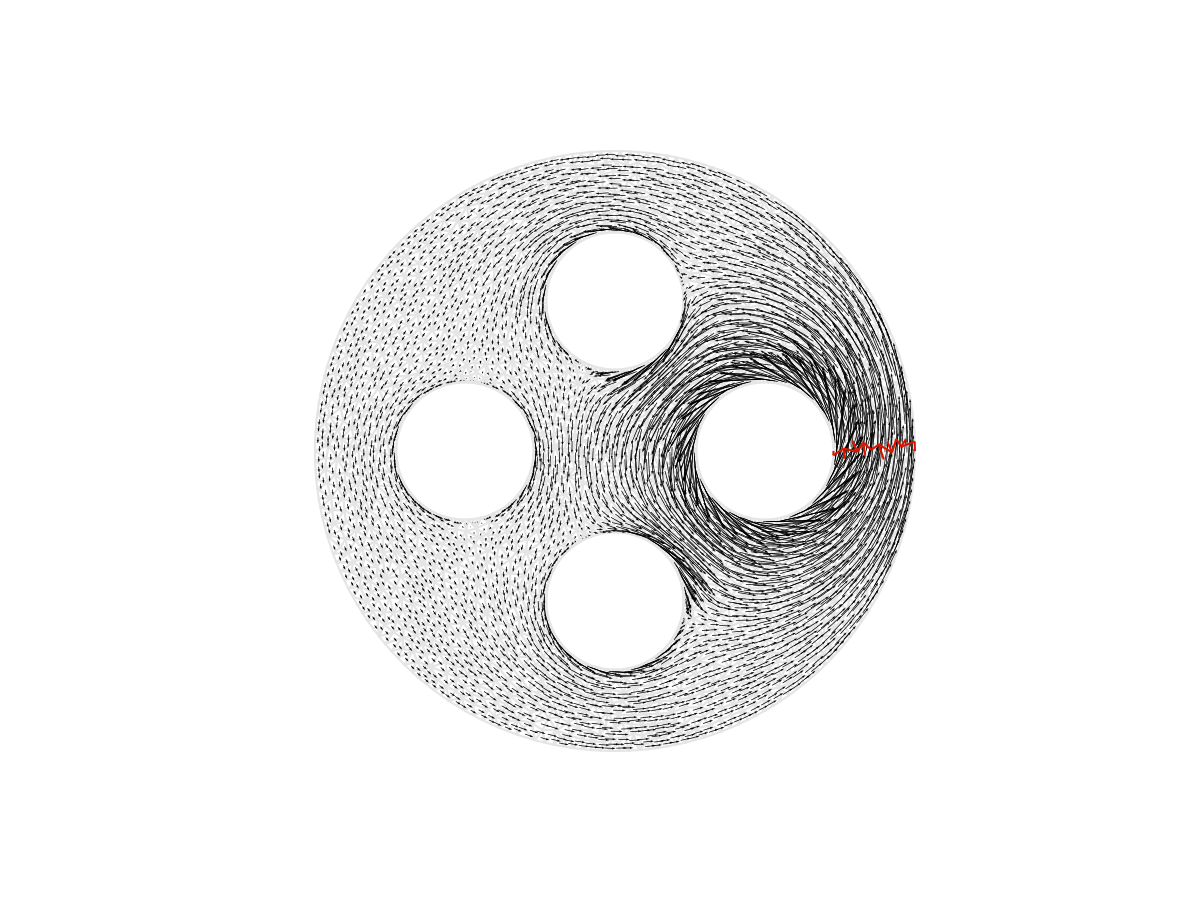} &
      \includegraphics[scale=0.35, trim=2.05in 0.95in 1.85in 0.95in,
      clip]
      {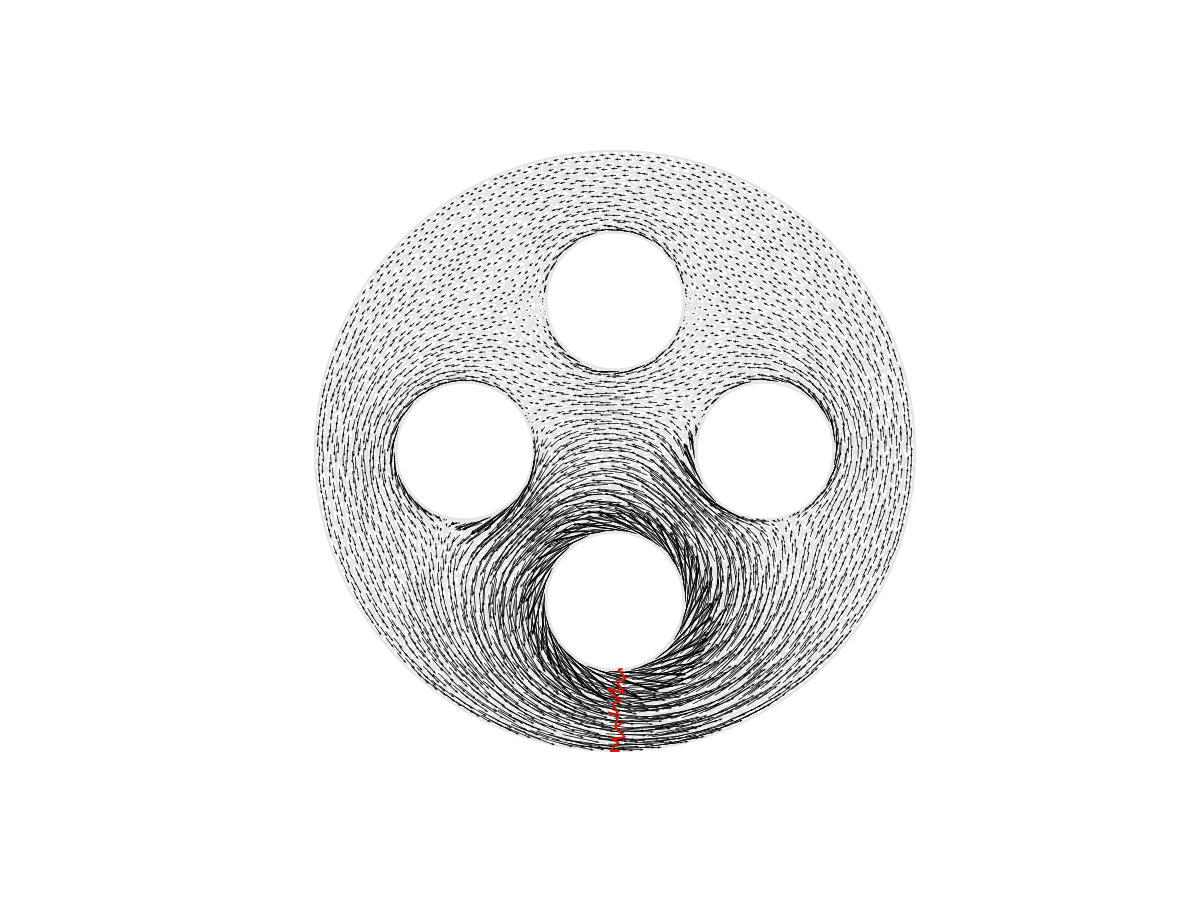} \\  
      \includegraphics[scale=0.35, trim=2.05in 0.95in 1.85in 0.95in,
      clip]
      {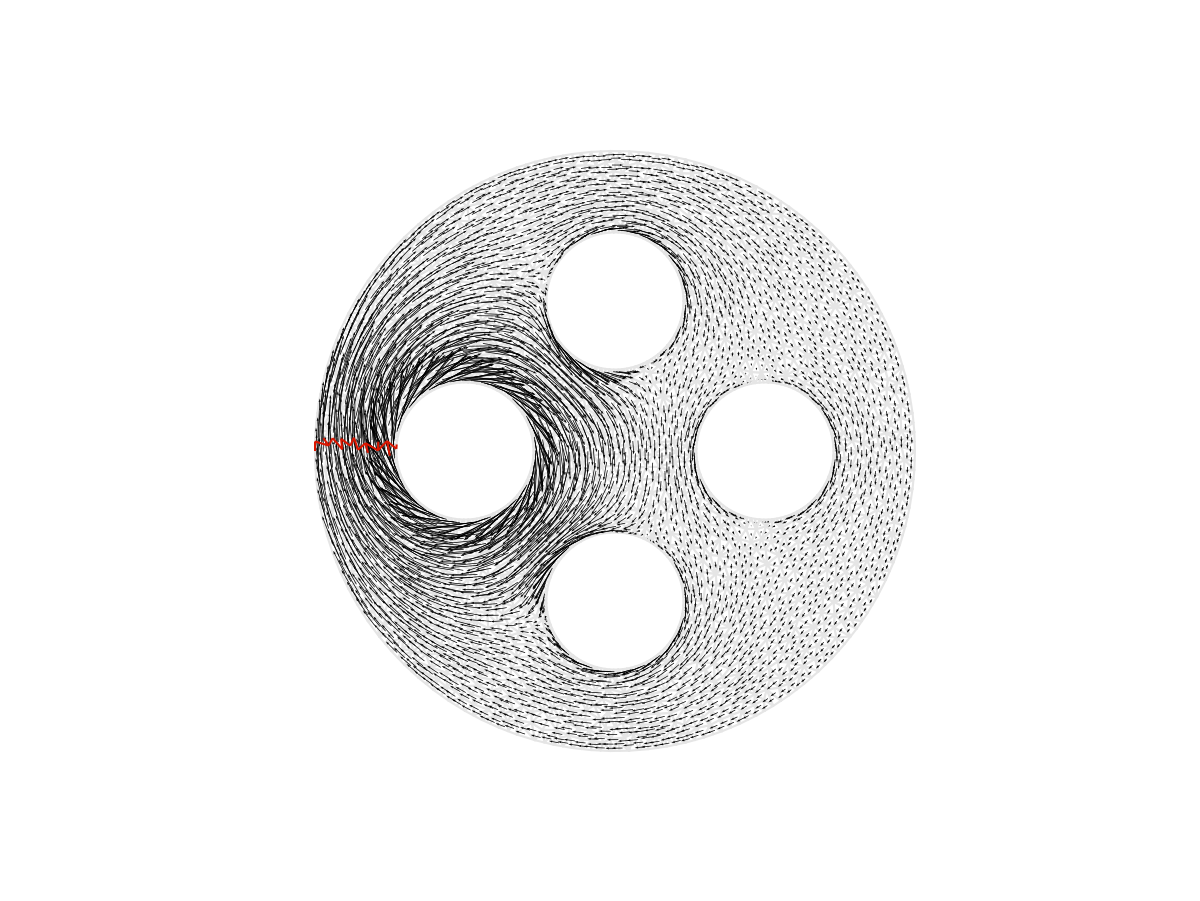} &
      \includegraphics[scale=0.35, trim=2.05in 0.95in 1.85in 0.95in,
      clip]
      {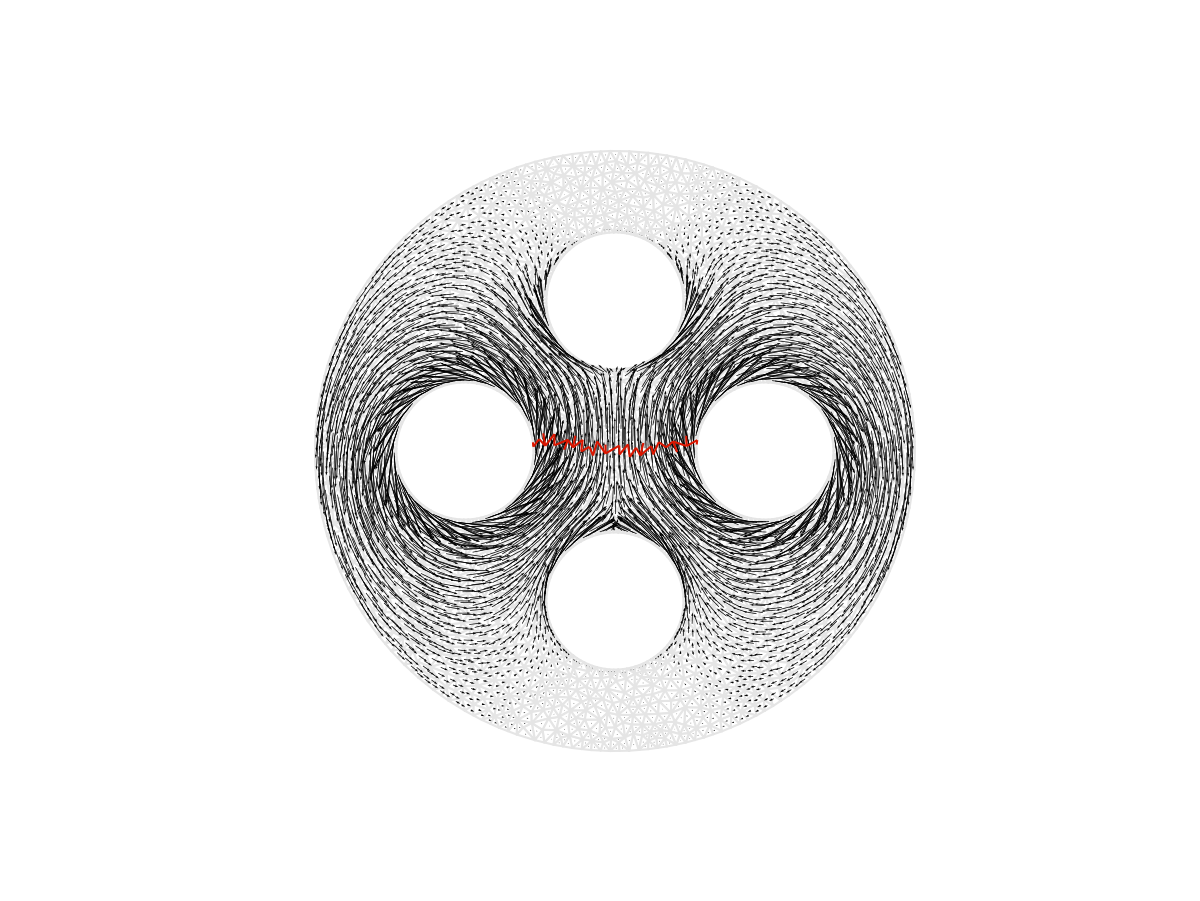} &
      \includegraphics[scale=0.35, trim=2.05in 0.95in 1.85in 0.95in,
      clip]
      {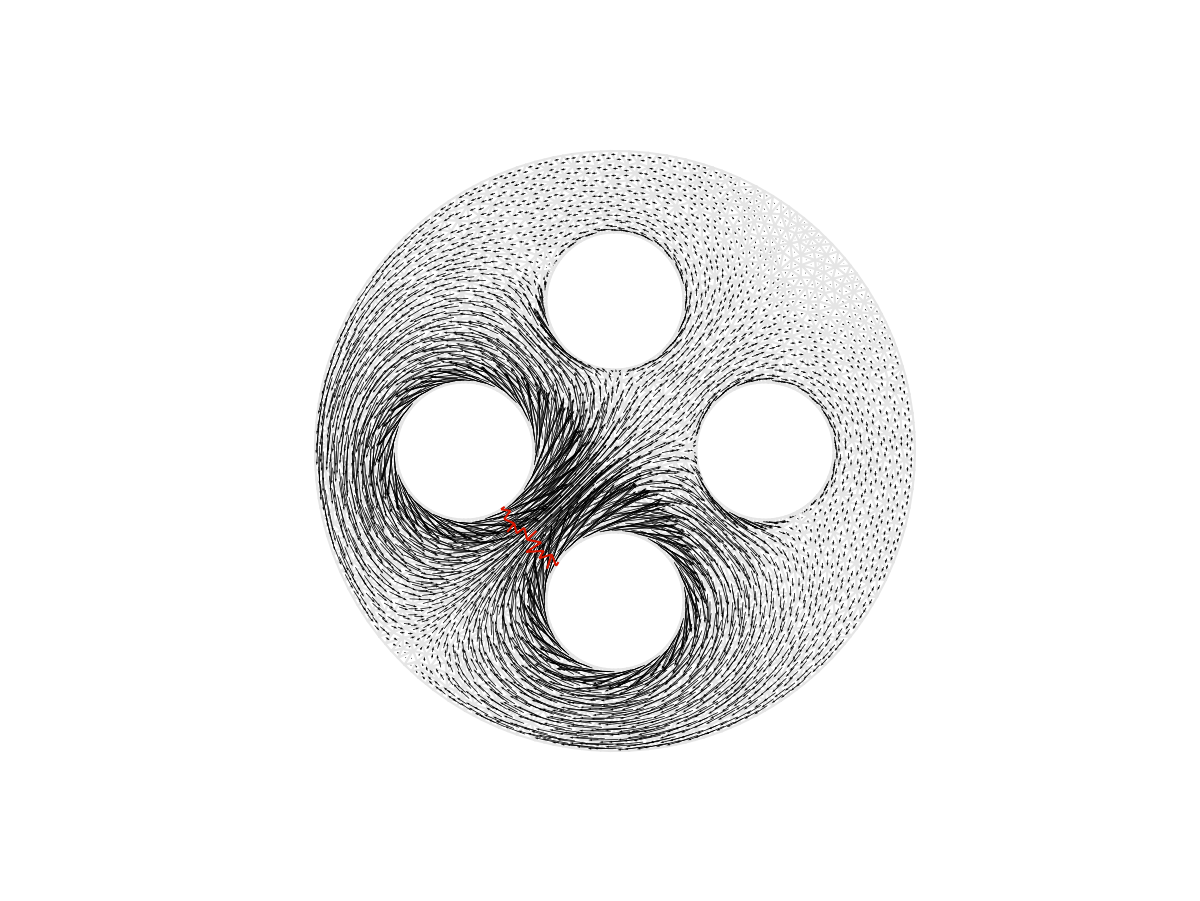}
    \end{tabular}
  \end{tabular}
  \caption{Some example computations using the least squares
    method. \emph{Top two rows~:} Cocycles representing a cohomology
    basis for the torus are shown as thick edges in the left
    figures. These are the $\omega$ cocycles of the text. The cocycles
    have value $\pm 1$ on these edges and 0 on the other edges. The
    right figures show the harmonic cochains in the corresponding
    cohomology classes. \emph{Bottom two rows :} The nontrivial
    cocycles are marked in red. Note that the proxy vector fields
    circulate \emph{around} only those holes associated with the
    cocycle, and \emph{past} others.}
  \label{fig:lstsqrschmlgy}
\end{figure}

\begin{figure}[t]
\centering
\begin{tabular}{ccc}
  \includegraphics[scale=0.35, trim=2.05in 0.95in 1.85in 0.95in, clip]
  {2d/fourholes_4326/illinois/3whtny.png}  &
  \includegraphics[scale=0.35, trim=2.05in 0.95in 1.85in 0.95in, clip]
  {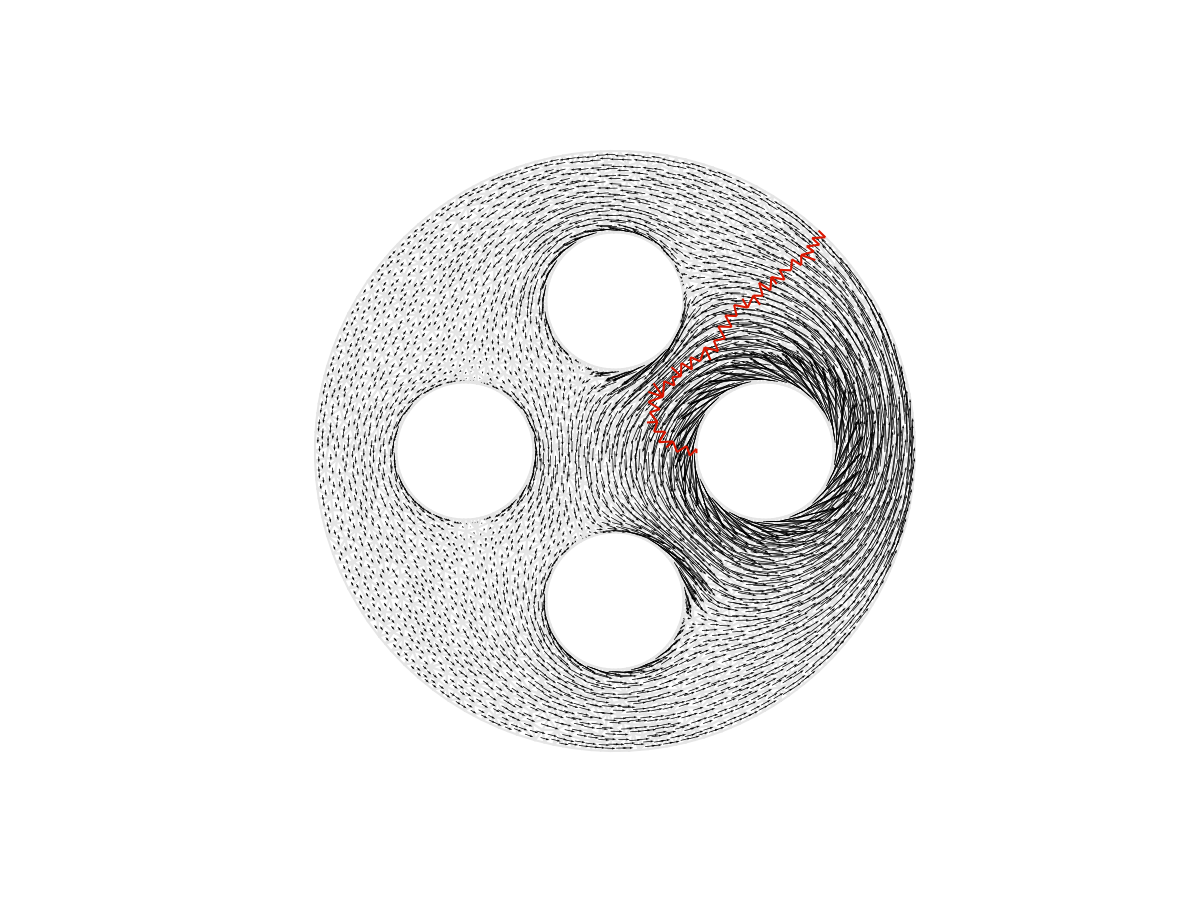} &
  \includegraphics[scale=0.35, trim=2.05in 0.95in 1.85in 0.95in, clip]
  {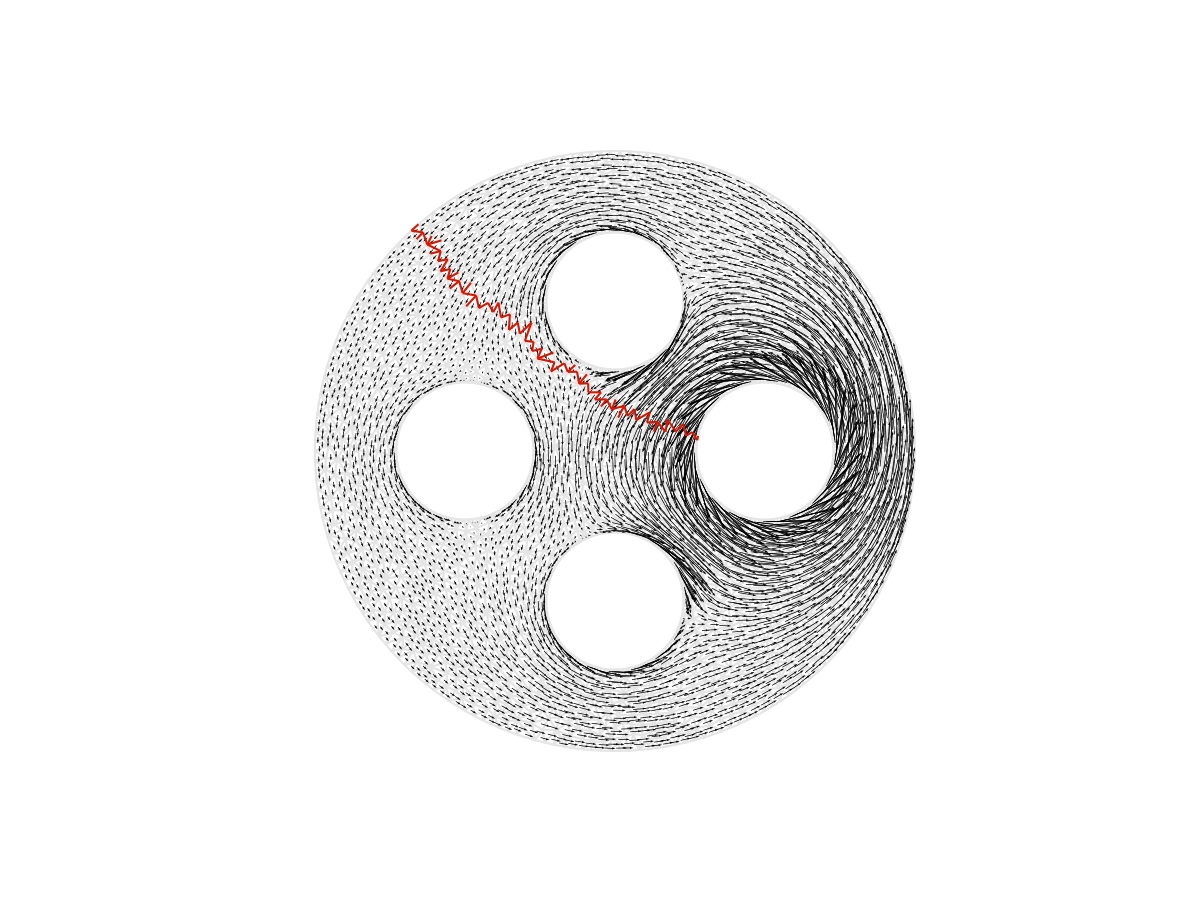} 
\end{tabular}
\caption{Three different cocycles ($\omega$ of the text) representing
  the same cohomology class lead to the same harmonic cochain when
  least squares method is used.}
\label{fig:2rprsnttvs}
\end{figure}

All computations in this paper were done using the Python language
with SciPy, NumPy, and PyDEC~\cite{BeHi2011a} packages.  The top two
rows of Figure~\ref{fig:lstsqrschmlgy} show the harmonic cochains
cohomologous to given nontrivial cocycles on a torus surface.  The
bottom two rows of Figure~\ref{fig:lstsqrschmlgy} show several
examples on a planar mesh with holes. To single out a particular hole,
so that the harmonic cochain proxy vector field will circulate around
that hole, one picks a cocycle connecting that boundary to the outer
boundary. Connecting two holes results in a harmonic cochain that
circulates about those two holes. For the cochains shown in the third
row of Figure~\ref{fig:lstsqrschmlgy}, from left to right, the values
of $\norm{\laplacian h}_{\hodge_1}$ relative to $\norm{h}_{\hodge_1}$
are approximately $4.12 \times 10^{-11}$, $7.32 \times 10^{-11}$ and
$3.02 \times 10^{-11}$, respectively. Similarly, for the cochains in
the bottom row, from left to right, these values are $4.98 \times
10^{-11}$, $5.73 \times 10^{-11}$ and $5.64 \times 10^{-11}$,
respectively.  Figure~\ref{fig:2rprsnttvs} shows that the least
squares method (as expected) finds the same harmonic cochain when very
different initial cocycles from the same cohomology class are given as
input. If the cohomologous cochains are denoted $h$, $h'$, $h''$ from
left to right, respectively, then the differences between them are
$\norm{h - h'}_{\hodge_1} = 1.1 \times 10^{-14}$, $\norm{h -
  h''}_{\hodge_1} = 2.8 \times 10^{-14}$ and $\norm{h' -
  h''}_{\hodge_1} = 2.2 \times 10^{-14}$.

\subsection{Linear solvers for the least squares
  method} \label{sec:solvers}

As noted earlier, the matrix $\d_{p-1}^T\hodge_p\d_{p-1}$
in~\eqref{eq:lstsqrs} is positive semidefinite since $\d_{p-1}$ will
typically have a nontrivial kernel. For example, for $p=1$ for a
connected domain, the space of constant functions on the domain is in
the kernel of $d_0$. In this case, it is easy to make the system
nonsingular (mod out the nontrivial kernel) by fixing the value at a
vertex and adjusting the linear system accordingly. For the case of
2-cochains in tetrahedral meshes however, the kernel of $\d_1$ can be
large. Let $M$ be a three-dimensional manifold simplicial
complex. Simple linear algebra and elementary topology reveals that
the $\dim(\ker \d_1) \ge N_0 - \chi(K)$ where $N_0$ is the number of
vertices and $\chi(K)$ is the Euler number (the alternating sum of
Betti numbers at all dimensions)~\cite{Munkres1984}. For example, for
a connected domain with boundary, we will have $\dim(\ker \d_1) \ge
\text{number of vertices} - 1 + \text{number of solid handles} -
\text{number of cavities}$. By refining the mesh this kernel dimension
can be made arbitrarily large. If a direct solver is to be used for
solving~\eqref{eq:lstsqrs} then one must mod out this potentially
large nontrivial kernel. An alternative is to use iterative Krylov
solvers as they work well even in the presence of a nontrivial kernel
and this is the approach we chose in our experiments. Specifically, we
used a conjugate gradient solver without any preconditioning or
modifications.  Algebraic multigrid is another very efficient
alternative whose effectiveness for this problem has been shown
in~\cite{Bell2008}.

\subsection{Finding the initial nontrivial
  cochains} \label{sub:initial}

In this paper we assume that a nontrivial cocycle is given. Our aim
here is not to give algorithms for finding a cocycle. However, a few
words about this are in order. An initial nontrivial cocycle in a
cohomology class can be found in a number of ways. For surfaces,
efficient algorithms to do this exist. By a folklore theorem, in time
linear in the number of simplices, one can find a homology basis for
the topological dual (e.g., barycentric dual) graph of the
triangulation. One can then use Poincar\'e-Lefschetz
duality~\cite{Munkres1984} to get a cohomology basis on the primal
mesh. For a boundaryless manifold simplicial complex, one would start
with nontrivial cycles on the dual graph. But in case of a manifold
with boundary, due to Lefschetz duality, one has to start with a
nontrivial relative cycle on the dual mesh, relative to the boundary.
One can also start with a random cochain and compute the desired
nontrivial cocycle using a Hodge decomposition with standard inner
product~\cite{Bell2008}. Yet another method is to use the persistence
algorithm~\cite{EdLeZo2002}. This is usually implemented using
coefficients in finite field $\F_2$ and has cubic (in the number of
simplices) complexity.

\section{Comparisons with Other Methods} \label{sec:comparisons}

The first relevant method to compare with is from the book of Gu and
Yau~\cite{GuYa2008} and also appears in their earlier work. The
formulation is very simple and straight forward, but it leads to
inefficient methods on general simplicial meshes. This method was
further simplified by Desbrun et al.~\cite{DeKaTo2008} who solve a
Poisson's-like equation at a different dimension. The resulting linear
systems in both methods suffer from numerical and scalability issues
for general simplicial meshes.

Gu and Yau start with a nontrivial cocycle $\omega$ representing a
cohomology class in $H^p(K)$ and seek a cochain $\omega + \d \alpha'$
such that $\laplacian (\omega+\d \alpha') = 0$. This leads to the
linear system $\d_{p-1}\hodge_{p-1}^{-1}\d_{p-1}^T\hodge_p\d_{p-1}
\alpha' = -\d_{p-1}\hodge_{p-1}^{-1}\d_{p-1}^T\hodge_p\omega$. The
presence of the inverse Hodge stars in this systems lead to numerical
disadvantages.

Desbrun et al.~\cite{DeKaTo2008} solve a different Poisson's equation
$(\d_{p - 2} \codiff_{p - 1} + \codiff_p \d_{p - 1}) \alpha'$ $= -
\codiff_p \omega$. A solution $\alpha'$ to the above equation yields
an $\omega+\d\alpha'$ that is harmonic in the sense of this paper. Of
course, if harmonic 1-cochains are being sought, then $\alpha'$ is a
0-cochain and $\codiff_0$ is the 0 operator. Thus the $\d \codiff$
term is not present. However, the $\d\codiff$ term is superfluous at
every dimension as we have shown. Thus their linear system has an
extra, unnecessary term. This extra term causes numerical and
scalability problems when Whitney Hodge star is used.

\begin{figure}[p]
  \centering
  \begin{tabular}{ccc}
    \imagetop{\begin{xyoverpic}
        {(1, 1)}{scale=0.4, trim=0.0in 0.0in 0.0in 0.0in, clip}
        {sprs/illnssldannls.png},
        (-0.05, 0.5)*{\begin{sideways}\text{Whitney}\end{sideways}},
        (0.5, 1.05)*{\text{Least squares matrix}} \end{xyoverpic}} &
    \imagetop{\begin{xyoverpic}
        {(1, 1)}{scale=0.4, trim=0.0in 0.0in 0.0in 0.0in, clip}
        {sprs/dsbrnsldannls.png},
        (0.5, 1.05)*{\text{Desbrun et al.~matrix}} \end{xyoverpic}} &
    \imagetop{\begin{xyoverpic}
        {(1, 1)}{scale=0.4, trim=0.0in 0.0in 0.0in 0.0in, clip}
        {sprs/dsbrnclrbr.png},
        (1.9, 0.98)*{10^{4}},
        (1.9, 0.84)*{10^{2}},
        (1.9, 0.685)*{10^{0}},
        (2.25, 0.53)*{10^{-2}},
        (2.25, 0.37)*{10^{-4}},
        (2.25, 0.215)*{10^{-6}},
        (2.25, 0.055)*{10^{-8}},
        (0.5, 1.08)*{\,}
      \end{xyoverpic}} \\
    \imagetop{\begin{xyoverpic}
        {(1,1)}{scale=0.4, trim=0.0in 0.0in 0.0in 0.0in, clip}
        {sprs/illnssldannlsdec.png},
        (-0.05, 0.5)*{\begin{sideways}\text{DEC}\end{sideways}}
      \end{xyoverpic}} &  
    \imagetop{\includegraphics[scale=0.4, trim=0.0in 0.0in 0.0in 0.0in, clip]
      {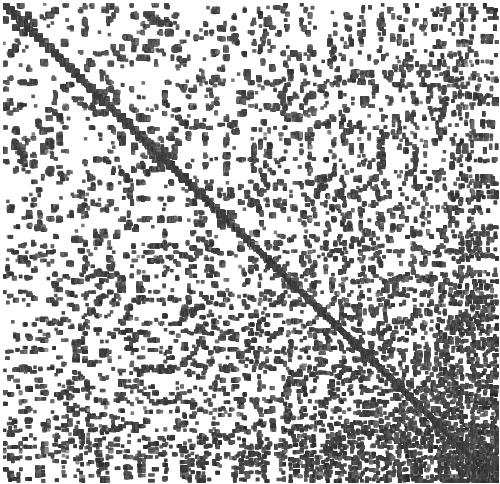}} & 
    \imagetop{\begin{xyoverpic}
        {(1, 1)}{scale=0.4, trim=0.0in 0.0in 0.0in 0.0in, clip}
        {sprs/dsbrnclrbrdec.png},
        (1.9, 0.98)*{10^{4}},
        (1.9, 0.84)*{10^{2}},
        (1.9, 0.685)*{10^{0}},
        (2.25, 0.53)*{10^{-2}},
        (2.25, 0.37)*{10^{-4}},
        (2.25, 0.215)*{10^{-6}},
        (2.25, 0.055)*{10^{-8}}
      \end{xyoverpic}}
  \end{tabular}
  \caption{Magnitudes of nonzeros in operators using the Whitney (top
    row) and DEC (bottom row) Hodge stars. The least squares matrix
    (left column) is sparser than the Desbrun et al.~matrix (right
    column) in the case of DEC Hodge star and significantly sparser in
    the case of Whitney Hodge star. This is due to the extra term in
    the Desbrun et al.~matrix. The colorbar shows the magnitude of the
    nonzero components. The two matrices are of equal size, and are
    for finding harmonic 2-cochains on the tetrahedral mesh of the
    solid annulus.}
  \label{fig:dnstydsbrn}
\end{figure}

In Figure~\ref{fig:dnstydsbrn}, we compare the sparsity of the least
squares and Desbrun et al.~matrices for finding harmonic 2-cochains on
a tetrahedral mesh of a solid annulus (a solid ball with an internal
cavity). The matrices are shown in Figure~\ref{fig:dnstydsbrn} for
both the Whitney and DEC Hodge stars. Both matrices are of the same
size but the Desbrun et al.~matrix is denser. This is very obvious for
the Whitney Hodge star case (14.3 million vs.~56 thousand
nonzeros). However, it is also evident in the DEC Hodge star case (94
thousand vs.~40 thousand nonzeros). Here the increased density is due
to the extra term in the Desbrun et al.~system.

The superior sparsity of the linear system matrix in the least squares
method leads to improved solution time. To illustrate this, we compare
the time taken for again finding harmonic 2-cochains on a tetrahedral
mesh of a solid annulus. For the least squares method, using conjugate
gradient method (without preconditioning), the times are 0.1355 and
0.1181 seconds for the DEC and Whitney Hodge stars, respectively. For
the Desbrun et al.~method, these times are 3.510 and 1746 seconds,
respectively. We also used a sparse solver in SuperLU for Desbrun et
al.~system and in this case, the times are 0.3171 and 13.05 seconds,
respectively. (All times are averaged over many trials. Also, it may
be possible to improve the times for both the methods by using
preconditioners or special solvers.) Another least square method is
that of Fisher et al.~\cite{FiScDeHo2007}. Comparisons with it are in
an earlier version of this paper available on
arXiv~\cite{HiKaWaWa2011v6}.

\section{Conclusions} \label{sec:cnclsns}

We presented two methods for finding harmonic cochains in the
cohomology class of a given cocycle -- an eigenvector method (using
direct or mixed formulation) followed by post processing and a least
squares method. The most salient feature of the least squares method
is in finding a cohomologous harmonic cochain without requiring an
entire harmonic or homology basis. The least squares method is
numerically superior and independent of the choice of Hodge stars in
comparison with the Poisson's equation methods of Gu and Yao, and
Desbrun et al. In future we plan to develop harmonic cochain methods
for higher order finite element exterior calculus analogous to the one
for Whitney forms. A precise quantification of the efficiency of the
least squares method in comparison with the eigenvector method for
finding a cohomologous harmonic basis is another direction to pursue.

\addcontentsline{toc}{section}{Acknowledgement}
\section*{Acknowledgement} This research was funded in part by NSF
Grant DMS-0645604. We thank Douglas Arnold, Alan Demlow, Tamal Dey,
Nathan Dunfield, Damrong Guoy, Rich Lehoucq, and Ari Stern for
discussions, and Mathieu Desbrun for pointing out the Fisher et
al.~paper.

\bibliographystyle{acmdoi} 
\addcontentsline{toc}{section}{References}
\bibliography{hirani}

\begin{thebibliography}{10}

\bibitem{AbMaRa1988}
{\sc Abraham, R., Marsden, J.~E., and Ratiu, T.}
\newblock {\em Manifolds, Tensor Analysis, and Applications}, second~ed.
\newblock Springer--Verlag, New York, 1988.

\bibitem{ArFaWi2010}
{\sc Arnold, D.~N., Falk, R.~S., and Winther, R.}
\newblock Finite element exterior calculus: from {H}odge theory to numerical
  stability.
\newblock {\em Bull. Amer. Math. Soc. (N.S.) 47}, 2 (2010), 281--354.
\newblock \href {http://dx.doi.org/10.1090/S0273-0979-10-01278-4} {\path{doi:
  10.1090/S0273-0979-10-01278-4}}.

\bibitem{BeHi2011a}
{\sc Bell, N., and Hirani, A.~N.}
\newblock {PyDEC}: Algorithms and software for {D}iscretization of {E}xterior
  {C}alculus, March 2011.
\newblock \href {http://arxiv.org/abs/1103.3076} {\path{arXiv:1103.3076}}.

\bibitem{Bell2008}
{\sc Bell, W.~N.}
\newblock {\em Algebraic Multigrid for Discrete Differential Forms}.
\newblock PhD thesis, University of Illinois at Urbana-Champaign, Urbana,
  Illinois, 2008.

\bibitem{Bjork1996}
{\sc Bj\"orck, A.}
\newblock {\em Numerical methods for least squares problems}.
\newblock Society for Industrial and Applied Mathematics (SIAM), Philadelphia,
  PA, 1996.

\bibitem{CaDeGlMi2005}
{\sc Cappell, S., DeTurck, D., Gluck, H., and Miller, E.~Y.}
\newblock Cohomology of harmonic forms on riemannian manifolds with boundary.
\newblock \href {http://arxiv.org/abs/0508372v1} {\path{arXiv:0508372v1}}.

\bibitem{DeKaTo2008}
{\sc Desbrun, M., Kanso, E., and Tong, Y.}
\newblock Discrete differential forms for computational modeling.
\newblock In {\em Discrete Differential Geometry}, A.~I. Bobenko, J.~M.
  Sullivan, P.~Schr\"oder, and G.~M. Ziegler, Eds., vol.~38 of {\em Oberwolfach
  Seminars}. Birkh\"auser Basel, 2008, pp.~287--324.
\newblock \href {http://dx.doi.org/10.1007/978-3-7643-8621-4_16} {\path{doi:
  10.1007/978-3-7643-8621-4_16}}.

\bibitem{EdLeZo2002}
{\sc Edelsbrunner, H., Letscher, D., and Zomorodian, A.}
\newblock Topological persistence and simplification.
\newblock {\em Discrete and Computational Geometry 28}, 4 (November 2002),
  511--533.
\newblock \href {http://dx.doi.org/10.1007/s00454-002-2885-2} {\path{doi:
  10.1007/s00454-002-2885-2}}.

\bibitem{FiScDeHo2007}
{\sc Fisher, M., Schr\"{o}der, P., Desbrun, M., and Hoppe, H.}
\newblock Design of tangent vector fields.
\newblock {\em ACM Transactions on Graphics 26}, 3 (July 2007), 56--1--56--9.

\bibitem{GuYa2008}
{\sc Gu, X.~D., and Yau, S.-T.}
\newblock {\em Computational Conformal Geometry}, vol.~3 of {\em Advanced
  Lectures in Mathematics (ALM)}.
\newblock International Press, Somerville, MA, 2008.

\bibitem{HiKaWaWa2011v6}
{\sc Hirani, A.~N., Kalyanaraman, K., Wang, H., and Watts, S.}
\newblock Cohomologous harmonic cochains, 2011.
\newblock Older longer version (version 6) of this paper.
\newblock \href {http://arxiv.org/abs/1012.2835v6} {\path{arXiv:1012.2835v6}}.

\bibitem{Jost2005}
{\sc Jost, J.}
\newblock {\em Riemannian Geometry and Geometric Analysis}, fourth~ed.
\newblock Universitext. Springer-Verlag, Berlin, 2005.
\newblock \href {http://dx.doi.org/10.1007/3-540-28890-2} {\path{doi:
  10.1007/3-540-28890-2}}.

\bibitem{Munkres1984}
{\sc Munkres, J.~R.}
\newblock {\em Elements of Algebraic Topology}.
\newblock Addison--Wesley Publishing Company, Menlo Park, 1984.

\bibitem{Schwarz1995}
{\sc Schwarz, G.}
\newblock {\em Hodge Decomposition---a Method for Solving Boundary Value
  Problems}, vol.~1607 of {\em Lecture Notes in Mathematics}.
\newblock Springer-Verlag, Berlin, 1995.

\end{thebibliography}

\end{document}